\documentclass[fleqn,usenatbib]{mnras}

\usepackage{newtxtext,newtxmath}
\usepackage[T1]{fontenc}
\usepackage{ae,aecompl}

\usepackage{graphicx}	
\usepackage{amsmath}	
\usepackage{amssymb}	

\usepackage[usenames,dvipsnames]{color}

\newcommand{\hl}{Hubble-Lema{\^i}tre}
\newcommand{\src}{em}
\newcommand{\obs}{obs}
\newcommand{\tsrc}{t_{\rm \src}}
\newcommand{\tobs}{t_{\rm \obs}}

\newcommand{\CIV}{C\,\textsc{iv}}

\newcommand{\FeII}{Fe\,\textsc{ii}}

\newcommand{\HI}{\textrm{H}\,\textsc{i}}

\newcommand{\Lya}{Ly$\alpha$}

\newcommand{\MgII}{\textrm{Mg}\,\textsc{ii}}
\newcommand{\NHI}{$N(\textrm{H}\,\textsc{i})$}

\newcommand{\SiII}{Si\,\textsc{ii}}

\newcommand{\SiIV}{Si\,\textsc{iv}}

\newcommand{\lcdm}{$\Lambda$CDM}
\newcommand{\OL}{$\Omega_{\Lambda}$}

\title[The ACCELERATION programme: I]{The ACCELERATION programme: I. Cosmology with the redshift drift\thanks{
Based on observations collected at the W.M. Keck Observatory which is operated as a scientific partnership among the California Institute of Technology, the University of California and the National Aeronautics and Space Administration. The Observatory was made possible by the generous financial support of the W.M. Keck Foundation.
}}

\author[R.~Cooke]{Ryan Cooke$^{1,2}$\thanks{email: ryan.j.cooke@durham.ac.uk}
\\
$^1$Centre for Extragalactic Astronomy, Department of Physics, Durham University, South Road, Durham DH1 3LE, UK\\
$^2$Royal Society University Research Fellow\\
}

\date{Accepted 6 Dec 2019. Received YYY; in original form ZZZ}

\pubyear{2019}

\begin{document}
\label{firstpage}
\pagerange{\pageref{firstpage}--\pageref{lastpage}}
\maketitle

\begin{abstract}
Detecting the change of a cosmological object's redshift due to the time evolution of the Universal expansion rate
is an ambitious experiment that will be attempted with future telescope facilities. In this paper, we
describe the ACCELERATION programme, which aims to study the properties of the most underdense regions of the
Universe. One of the highlight goals of this programme is to prepare for the redshift drift measurement.
Using the EAGLE cosmological hydrodynamic simulations, we estimate the peculiar acceleration of gas in galaxies and
in the \Lya\ forest. We find that star-forming `cold neutral gas' exhibits large peculiar acceleration
due to the high local density of baryons near star-forming regions. We conclude that absorption by cold neutral gas is unlikely to
yield a detection of the cosmological redshift drift. On the other hand, we find that the peculiar accelerations of \Lya\ forest absorbers are more than an order of magnitude below the expected cosmological signal. We also highlight that
the numerous low \HI\ column density systems display lower peculiar acceleration. Finally, we propose a new
`\Lya\ cell' technique that applies a small correction to the wavelength calibration to secure a relative measurement
of the cosmic drift between two unrelated cosmological sources at different redshifts. For suitable combinations of
absorption lines, the cosmological signal can be more than doubled, while the affect of the observer peculiar
acceleration is mitigated. Using current data of four suitable \Lya\ cells, we infer a limit on the cosmological
redshift drift to be $\dot{v}_{\rm obs}<65~{\rm m~s}^{-1}~{\rm year}^{-1}$ ($2\sigma$).
\end{abstract}


\section{Introduction}

We have now been studying the cosmological motion of matter in the Universe for almost a century \citep{Lem27,Hub29}, and learned that matter on cosmological scales appears to recede from us. However, we do not understand the origin of this Universal expansion. Furthermore, modern measurements of standard rulers and candles have shown us that the expansion is accelerating, which can either be due to a problem with gravity on cosmological scales or the existence of some unknown content generally referred to as `dark energy'. Given these unanticipated discoveries, and our current inability to explain these phenomena, it is imperative that we continue our efforts to understand the cosmological dynamics of matter in the Universe.

To date, all cosmological observations rely on a model (for example, cosmological distances are not \emph{directly} measured, they are modelled); we currently lack a \emph{direct} detection of the cosmological motion of matter in the Universe. In theory, the Universal expansion causes objects to change their brightness and redshift. However, since a human lifetime is considerably less than the age of the Universe, a measurable change in either the apparent luminosity or redshift is comparatively small. The idea of `watching the Universe expand in real time' was first envisioned by \citet[][see also \citealt{McV62}, who derived the equations for a Universe with a cosmological constant]{San62}. Unlike all other cosmological experiments currently conceived, this is the only experiment where the signal will continue to grow larger with time. Further to this point, provided there is a patient enough observer, this experiment will (at least in principle) always overcome any systematic uncertainties, given sufficient time! Considering the difficulty of obtaining high precision measurements of the apparent magnitude of cosmological sources (i.e. due to signal-to-noise limitations and from the intrinsic luminosity variations of astrophysical sources), experiments of the so-called `Sandage Test' have instead focused on the redshift drift.

\begin{figure}
	\includegraphics[width=\columnwidth]{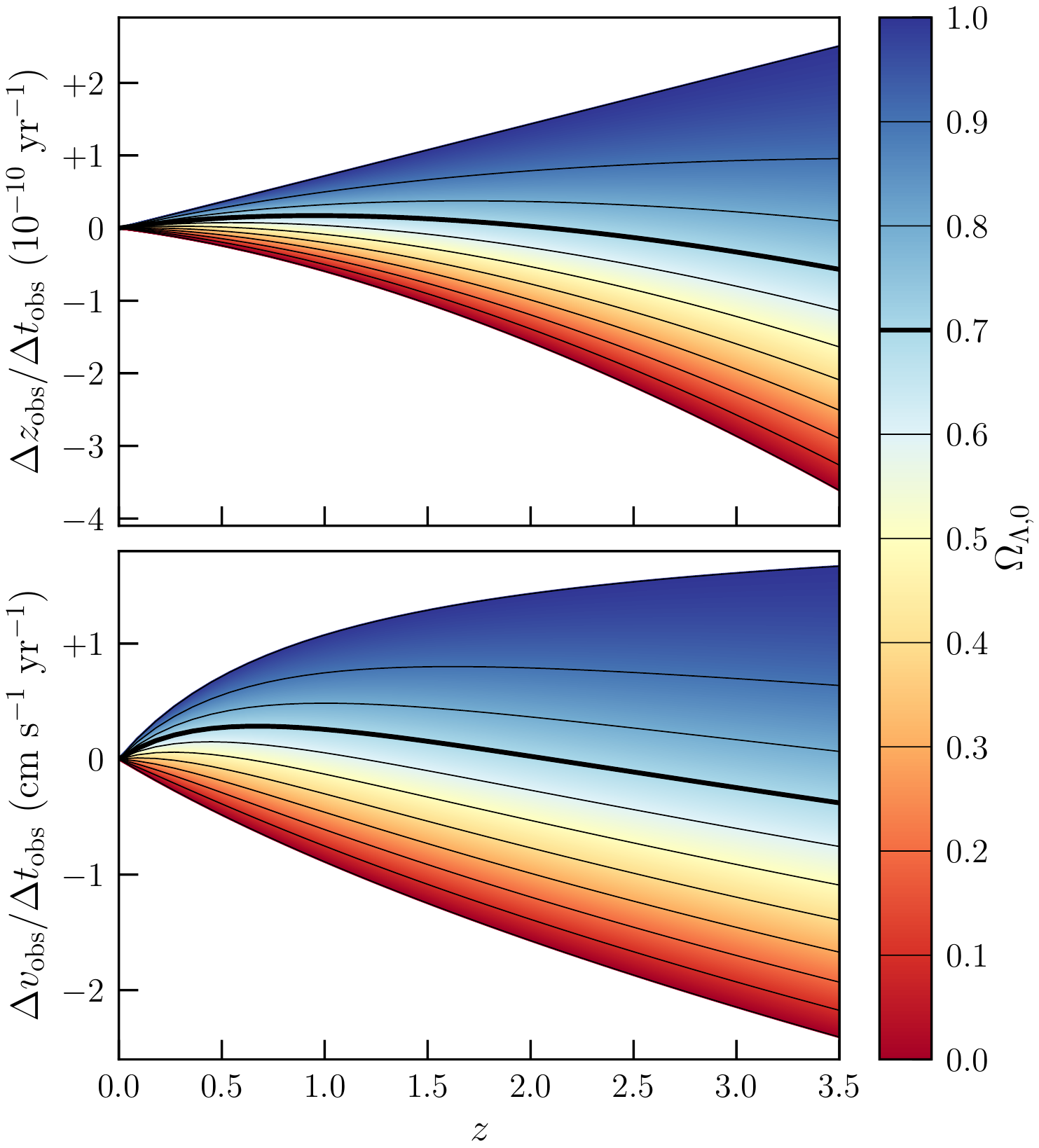}
    \caption{The top panel shows the redshift drift,
    $\dot{z}_{\rm obs}\equiv\Delta z_{\rm obs}/\Delta t_{\rm obs}$
    as a function of redshift, colour-coded by the value of \OL\ in a flat Universe. The black contour lines show the redshift drift for several representative values of \OL, as indicated on the colourbar. The thick black line approximately represents the standard \lcdm\ curve. The bottom panel displays the velocity drift,
    $\dot{v}_{\rm obs}=\Delta v_{\rm obs}/\Delta t_{\rm obs}= c\dot{z}_{\rm obs}/(1+z)$ as a function of redshift, with all curves having the same meaning as the top panel.
    }
    \label{fig:accel}
\end{figure}

A measurement of the redshift drift requires a `simple' monitoring experiment of the position of spectral lines over multiple epochs. Photons that are emitted at times $\tsrc$ and $\tsrc+\Delta \tsrc$ by a comoving source will be received at times $\tobs$ and $\tobs+\Delta\tobs$, respectively, by a comoving observer (i.e. assuming that both the observer and source have no peculiar velocity or acceleration; for a more detailed derivation, see \citealt{Lis08}). In this case, the redshift drift can be calculated as follows:
\begin{equation}
\dot{z}_{\rm \obs}\Delta\tobs=\Delta z_{\rm \obs} = \frac{a(\tobs+\Delta\tobs)}{a(\tsrc+\Delta\tsrc)} - \frac{a(\tobs)}{a(\tsrc)}
\end{equation}
To first order, $a(\tsrc+\Delta\tsrc) \approx  a(\tsrc)+\dot{a}(\tsrc)\Delta\tsrc$, and recall that $\Delta\tsrc=\Delta\tobs a(\tsrc)/a(\tobs)$. Thus, the redshift drift is given by
\begin{equation}
\label{eqn:dzdt}
\dot{z}_{\rm \obs}\equiv\frac{\Delta z_{\rm \obs}}{\Delta t_{\rm \obs}} \simeq (1+z_{\rm obs})H_{0} - H(z_{\rm obs})
\end{equation}
where $H(z_{\rm obs})$ is the \hl\ parameter at redshift $z_{\rm obs}$. The redshift drift and the corresponding velocity drift, $\dot{v}_{\rm \obs}\equiv c\dot{z}_{\rm \obs}/(1+z)$, are shown in the top and bottom panels of Figure~\ref{fig:accel} respectively, where the thick black line corresponds to a flat $\Lambda$CDM cosmological model, with parameters based on the \citet{Pla18} results. The typical spectral shift of a cosmological source in one year is expected to be a few mm/s.

Although this number sounds like a challenging goal, many groups have made projections for the next generation of telescope facilities in order to realize a direct measurement of cosmic dynamics. A promising approach, first proposed by \citet{Loe98}, utilizes the forest of \HI\ Lyman-$\alpha$ absorption lines that are imprinted on the spectrum of a background $z\sim3$ quasar.\footnote{This approach is sometimes termed the `Sandage-Loeb' test, which specifically refers to the Sandage test applied to the Ly$\alpha$ forest.} Indeed, the redshift drift experiment became one of the defining goals of the ESO Extremely Large Telescope (ELT). The first detailed feasibility study for \Lya\ forest measurements with the ELT was conducted by \citet{Lis08}. These authors conclude, like \citet{Loe98}, that the \Lya\ forest should be a relatively clean probe of cosmic dynamics, and will require a few thousand hours of observing time over a time baseline of a few decades.

An alternative approach, considered briefly by \citet{Lis08}, is to measure the redshift drift of either molecular absorption lines or the \HI\ 21~cm absorption line imprinted on sub-mm/radio-bright quasars. This approach has the advantage that the lines are intrinsically narrow, which allows the line centre to be more precisely measured. Indeed, the current best constraint on the redshift drift utilizes this approach \citep{Dar12}. These authors also predict that this approach could yield a direct measure of cosmic acceleration with the Square Kilometer Array (SKA) with baselines of just $\sim5$ years. These estimates require several hundreds of absorption line sources to be discovered, which may be leveraged by discoveries with the Five hundred meter Aperture Spherical Telescope (FAST; \citealt{Jiao19}).\footnote{\citet{Jiao19} also performed a pilot study to measure the redshift drift and found a $\sim3~{\rm km~s}^{-1}$ shift which is attributed to an unknown systematic.}

SKA may also be able to detect the redshift drift using \HI\ 21~cm \emph{emission} from galaxies out to a redshift $z\sim1$ \citep{Klo15}. Although this seems unfeasible with SKA Phase I, it may become possible with just $\sim12$ years of observations with Phase II of the SKA, provided that some modifications are made to the SKA baseline specification. More recently, \citet{Alv19} presented a forecast of the redshift drift from the combination of \Lya\ forest, SKA, and \HI\ intensity mapping experiments. These authors highlight the importance of using probes covering a range of redshift to break degeneracies and improve the precision with which cosmological parameters can be estimated.

\subsection{Instrumental Systematics}

With a view to overcome the difficulties associated with long time baseline measurements, \citet{Kim15} describe alternative approaches to the redshift drift measurement, with a focus on low redshift measurements. They also consider different instrument designs that might alleviate the dominant instrument systematics. Overall, instrument-related systematics are currently the main limitation of the redshift drift experiment. Foremost, all of the above techniques require a stable spectrograph, and an Earth model that includes an accurate barycentric correction \citep[e.g.][]{LinDra03}.

High redshift \Lya\ forest measurements use optical echelle spectrographs, which are known to suffer systematic uncertainties at the level of a few $100~{\rm m~s}^{-1}$ \citep{WhiMurGri10,Rah13,WhiMur15} when using standard Thorium-Argon calibration procedures. The current gold standard in optical wavelength calibration is at the level of $\sim0.5~{\rm m~s}^{-1}$ \citep[e.g.][]{Dum12,Cer19,Cof19} using laser frequency comb technology \citep{Mur07,Wil12}. In the immediate future, the Echelle SPectrograph for Rocky Exoplanet and Stable Spectroscopic Observations (ESPRESSO) hopes to achieve a wavelength precision and stability of $\sim10~{\rm cm~s}^{-1}$, also using laser frequency combs.

\begin{figure}
	\includegraphics[width=\columnwidth]{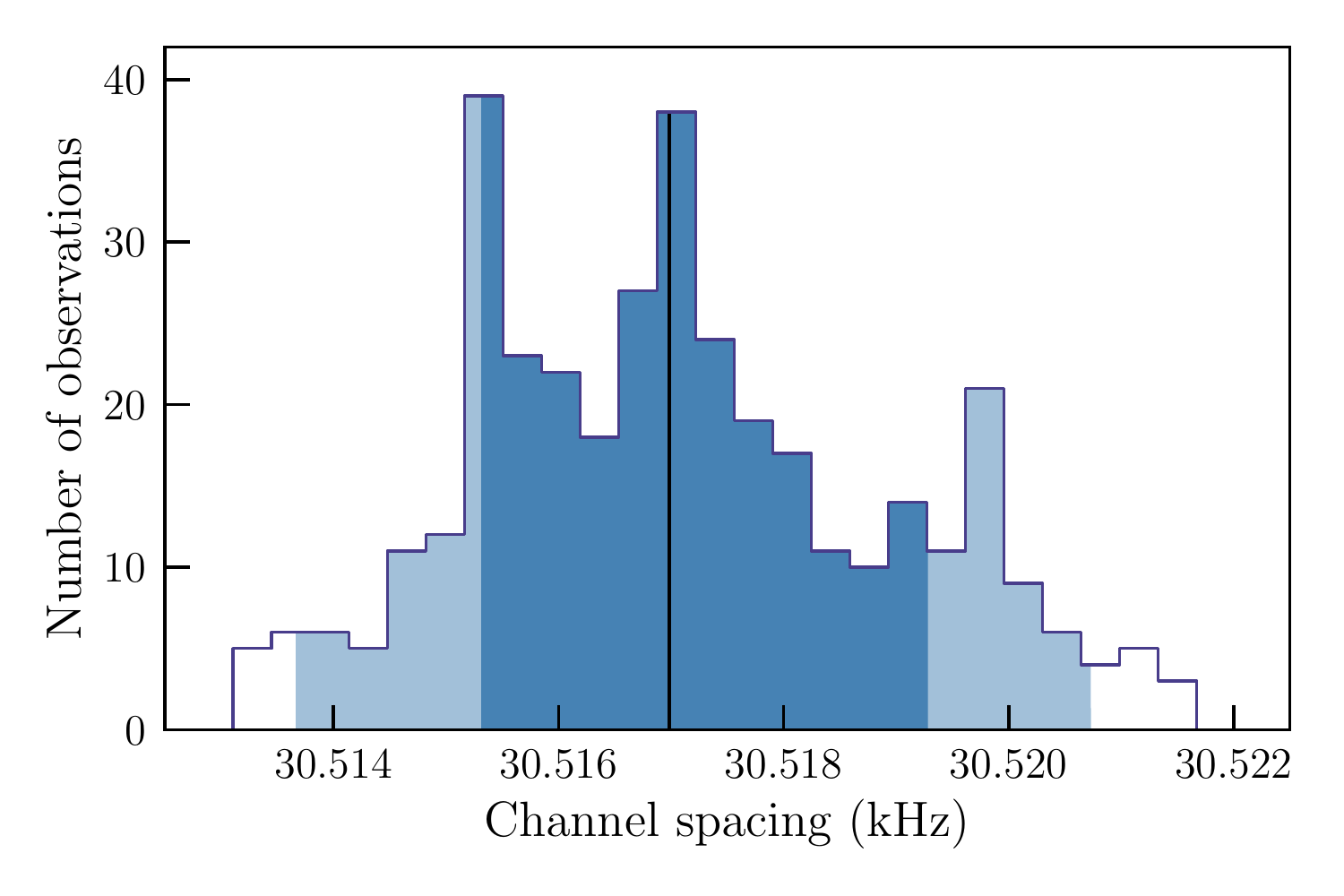}
    \caption{The distribution of channel spacing for all high spectral resolution ALMA data acquired during the period 2011--2018. The dark and light shaded regions correspond to the inner 68 and 95 percent of the observations, respectively. The vertical line indicates the median value. This distribution indicates that 68 percent of the observations have an identical channel width to within 1 part in 8000 ($\simeq0.5~{\rm cm~s}^{-1}$) over a period of 7 years. For comparison, the cosmological signal is an order of magnitude \emph{larger} than the precision reached with ALMA.
    }
    \label{fig:chanspace}
\end{figure}

Further improvement on the wavelength/frequency stability is possible in the sub-mm/radio domain, which uses heterodyne instrumentation. Using the Atacama Large Millimeter Array (ALMA) as an example, the primary limitation of the frequency stability is the Local Oscillator, the precision of which can be controlled to much better than the expected cosmological signal, especially at high spectral resolution. To demonstrate the advantage of this instrumentation, one of the highest ALMA frequency resolution settings ($\sim30.5~{\rm kHz}\simeq35~{\rm m/s}$ channel spacing at 250~GHz), corresponds to a lag of $33\,\mu{\rm s}$. Since the timing standards are accurate to within a few nanoseconds, the channel labelling is accurate to within $\sim1/10000$ of a frequency channel (i.e. $\sim0.3~{\rm cm~s}^{-1}$); the accuracy is even more improved when using a higher spectral resolution mode. Similarly, the width of a frequency channel is well-determined. The distribution of channel spacing based on archival ALMA data\footnote{Available from: \url{http://almascience.nrao.edu/aq}} (from 2011--2018) is shown in Figure~\ref{fig:chanspace}; this distribution indicates that 68 percent of the observations have an identical channel width to within 1 part in 8000 (corresponding to a velocity precision of $\simeq0.5~{\rm cm~s}^{-1}$) over a period of seven years; recall the cosmological signal is roughly an order of magnitude greater than this limit over the same seven year baseline.

\subsection{The ACCELERATION programme}

The overarching goal of the ACCELERATION\footnote{ACCELERATION: Automated Convnet and Chi-squared Estimation of the Late-time Expansion Rate of Absorbers, Temperature of the IGM, and the Observed \NHI\ distribution.} programme is to characterize the forest of absorption lines that are imprinted on optical quasar spectra, primarily to study the properties of the most underdense regions of the Universe.

This paper presents the foundations for one of the main objectives of the ACCELERATION programme: To work towards a measurement of the cosmological redshift drift. In Section~\ref{sec:sim}, we employ a recent suite of cosmological hydrodynamic simulations to reassess the peculiar acceleration of sources, and discuss the instrumental and observational systematics that might inhibit a measurement of the redshift drift. In Section~\ref{sec:obs}, we introduce a new technique that overcomes several instrumental and observational systematics. We then leverage data of a well-observed high redshift quasar to place a limit on the cosmological redshift drift using the \Lya\ forest. A summary of the main conclusions are listed in Section~\ref{sec:conc}.

\section{Peculiar Accelerations}
\label{sec:sim}

To secure a measure of the cosmic redshift drift, we must correct for the peculiar acceleration of both the source and the observer. The acceleration of the solar system barycentre is now known to high precision using very long baseline interferometer observations of extragalactic radio sources \citep{TitKra18}, and this is expected to improve with the final \emph{Gaia} data release \citep{Gaia18}. Using a recent determination of the distance to the Galactic centre and the Galaxy rotation velocity at the Sun's galactocentric distance \citep{deGBon16}, the magnitude of the acceleration vector of the solar system barycentre is $|a_{\odot}| \simeq 0.76\pm0.03~{\rm cm~s}^{-1}~{\rm year}^{-1}$, in the direction of the Galactic centre. Note, the solar system acceleration is a dipole, while the cosmic signal is expected to be uniform in all directions at a given redshift. Related to this point, the magnitude of the barycentric acceleration is $\approx0$ when observing extragalactic sources that are orthogonal to the vector pointing towards the Galactic centre. Later, in Section~\ref{sec:obs}, we will also discuss an observational strategy that can eliminate the uncertainty of the observer reference frame. In summary, the motion of the Earth can be fully accounted for to sufficient precision to detect the cosmological redshift drift.

\begin{figure*}
	\includegraphics[width=\textwidth]{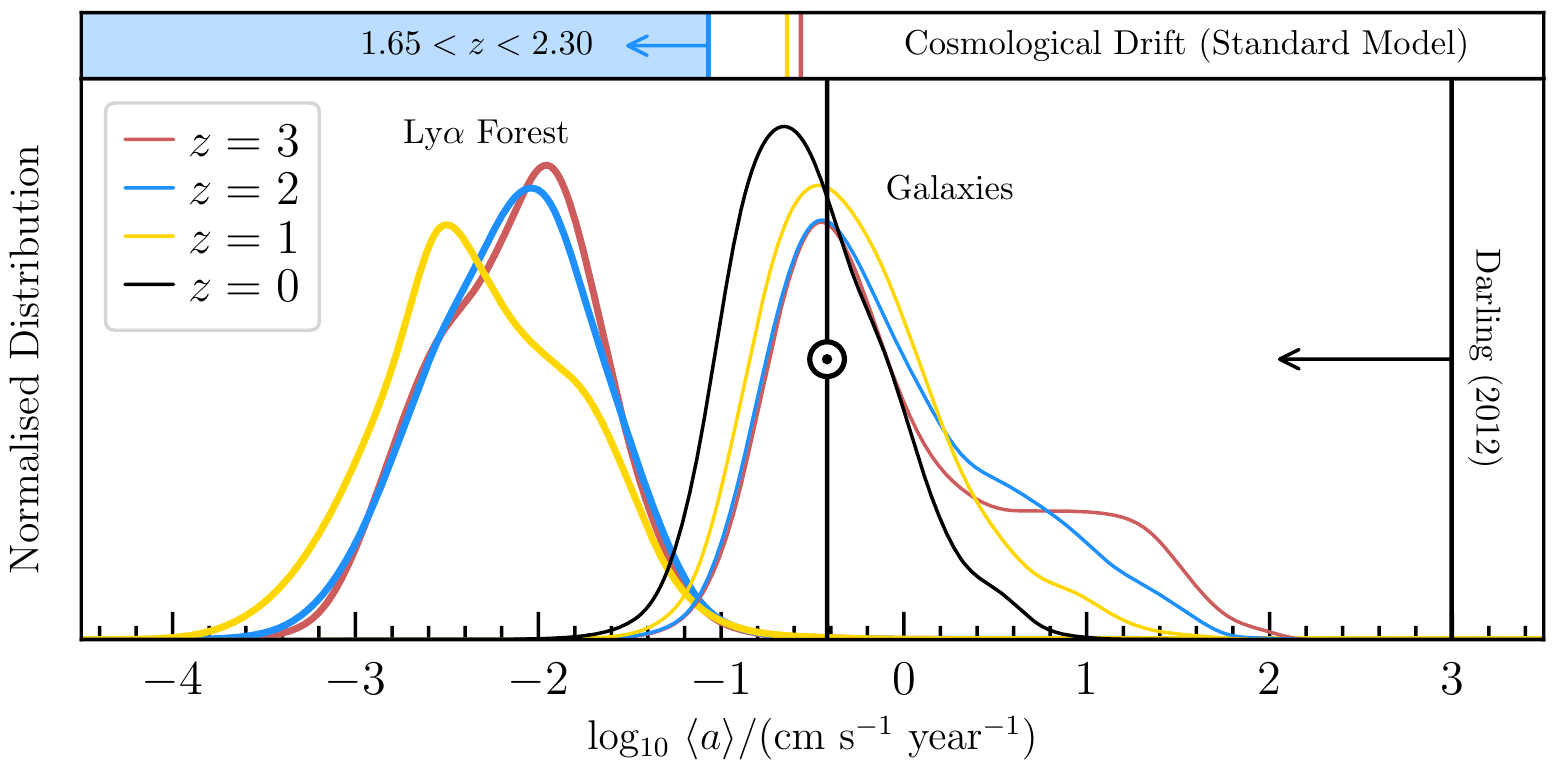}
    \caption{The main panel shows the probability distributions, ${\rm Pr}(\langle a \rangle)$, of the angle-averaged acceleration of star-forming gas in galaxies (thin curves) and the \Lya\ forest (thick curves). The black, yellow, blue, and red curves represent the distributions at redshift $z=0, 1, 2, 3$, respectively. The current $2\sigma$ limit on cosmic acceleration reported by \citet{Dar12} is shown by the vertical line and horizontal arrow on the right of the plot, $\log_{10}\langle a\rangle/({\rm cm~s}^{-1}~{\rm year}^{-1})\le3$. The angle-averaged acceleration of the Sun is shown by the vertical line with the $\odot$ symbol (for reference, this value should be multiplied by 2 to get the magnitude of the solar acceleration). The coloured vertical lines in the top panel indicate the expected cosmic velocity drift for the Standard \lcdm\ model (with the same colour-coding as the main panel). Note, at redshift $1.65<z<2.30$, the cosmological velocity drift (for the concordant \lcdm\ model, as seen by an observer at $z=0$) is comparable to the peculiar acceleration of the \Lya\ forest.}
    \label{fig:sim}
\end{figure*}

The peculiar acceleration of external galaxies and galaxy clusters has been investigated by many authors \citep{Phi82,Lak82,Teu86,Ame08}. The general conclusion from these works is that the peculiar acceleration of galaxies can be mitigated by observing many independent sources. To give a ballpark estimate, we might expect the peculiar acceleration of gas in external galaxies to be of similar magnitude to the angle-averaged acceleration of the solar system barycentre (i.e. $\langle a \rangle = |a_{\odot}|/2 \simeq 0.4~{\rm cm~s}^{-1}~{\rm year}^{-1}$), which is comparable to the cosmological value. However, by observing $N$ random and uncorrelated sources, the net uncertainty of the source peculiar acceleration decreases by $\sqrt{N}$.

The expected peculiar acceleration of the \Lya\ forest has also been investigated \citep[e.g.][]{Loe98,Lis08}. Both of these authors conclude that the peculiar acceleration of the \Lya\ forest is several orders of magnitude below the cosmological signal, which is an attractive property of using the \Lya\ forest to secure a measurement of cosmic acceleration.

In this section, we use the EAGLE\footnote{Evolution and Assembly of GaLaxies and their Environments (EAGLE), available from: \url{http://icc.dur.ac.uk/Eagle/}} cosmological hydrodynamic simulations \citep{Cra15,Sch15} to calculate the peculiar acceleration of gas in galaxies and the \Lya\ forest across a range of redshift. In order to obtain good statistics, we use the Ref-L100N1504 simulation, which is 100~cMpc on a side and contains $1504^{3}$ dark matter particles of mass $m_{\rm dm}=9.7\times10^{6}~M_{\odot}$. The simulation was calculated assuming a flat $\Lambda$CDM model with the following cosmological parameters: $H_{0}=67.8~{\rm km~s}^{-1}~{\rm Mpc}^{-1}$, $\Omega_{\Lambda}=0.693$, $\Omega_{\rm B,0}\,h^{2}=0.02216$. The main goal of this paper is to estimate the peculiar acceleration of cold gas in galaxies (i.e. gas that might be observed as molecular or \HI\ absorption in front of a sub-mm/radio bright quasar) or low density gas in the intergalactic medium (i.e. gas in front of an optically bright quasar).

Throughout this paper, we calculate the strength of the local gravitational field at a position $\Vec{r_{0}}$, and use this as a proxy of the peculiar acceleration:
\begin{equation}
    \Vec{a} = -\sum \frac{G\,m_{\rm part,i}\,(\Vec{r_{\rm i}}-\Vec{r_{0}})}{\vert\Vec{r_{\rm i}}-\Vec{r_{0}}\vert^3}
\end{equation}
where $G$ is the Newton gravitational constant, and the vector sum is over all particles (including star, gas, dark matter, and black hole particles) of mass $m_{\rm part,i}$ and distance $\Vec{r_{\rm i}}-\Vec{r_{0}}$ from the position of interest. We always report an angle-averaged acceleration:
\begin{equation}
    \langle a \rangle = \int_{0}^{\pi/2} \vert \Vec{a}\vert\,\cos{\theta}\,\sin{\theta}\,{\rm d}\theta = \vert\Vec{a}\vert/2
\end{equation}
where $\theta$ is the angle between the acceleration vector and the line of sight. (i.e. we report the magnitude of the acceleration divided by two).

\subsection{The peculiar acceleration of gas in galaxies}
\label{sec:simcnm}

EAGLE does not explicitly simulate cold molecular gas, but we can nevertheless estimate the gravitational field in the environments that might host cold molecular gas. We first identify gas particles that are converted into star particles in an adjacent simulation output, and calculate the total gravitational field strength at the location of the gas particle. To minimize the time between simulation outputs, we use the particle data from the snipshots, which have a time difference of $\simeq38.2,~55.4,~98.9,~57.5~{\rm Myr}$ in the $z=3,~2,~1,~0$ outputs, respectively.\footnote{We explicitly checked that the results are unchanged for the $z=3$ output by using snipshots with a time difference of $\simeq75.8~{\rm Myr}$.} To speed up the calculation for this test, only particles with the same \texttt{GroupNumber} are included in the calculation of the local gravitational field.\footnote{The \texttt{GroupNumber} is an integer that labels individual particles to a given Friends-of-Friends halo.}
The angle-averaged acceleration distributions of the four representative redshift values are shown in Figure~\ref{fig:sim}, as the thin coloured curves.

\begin{figure*}
	\includegraphics[width=\textwidth]{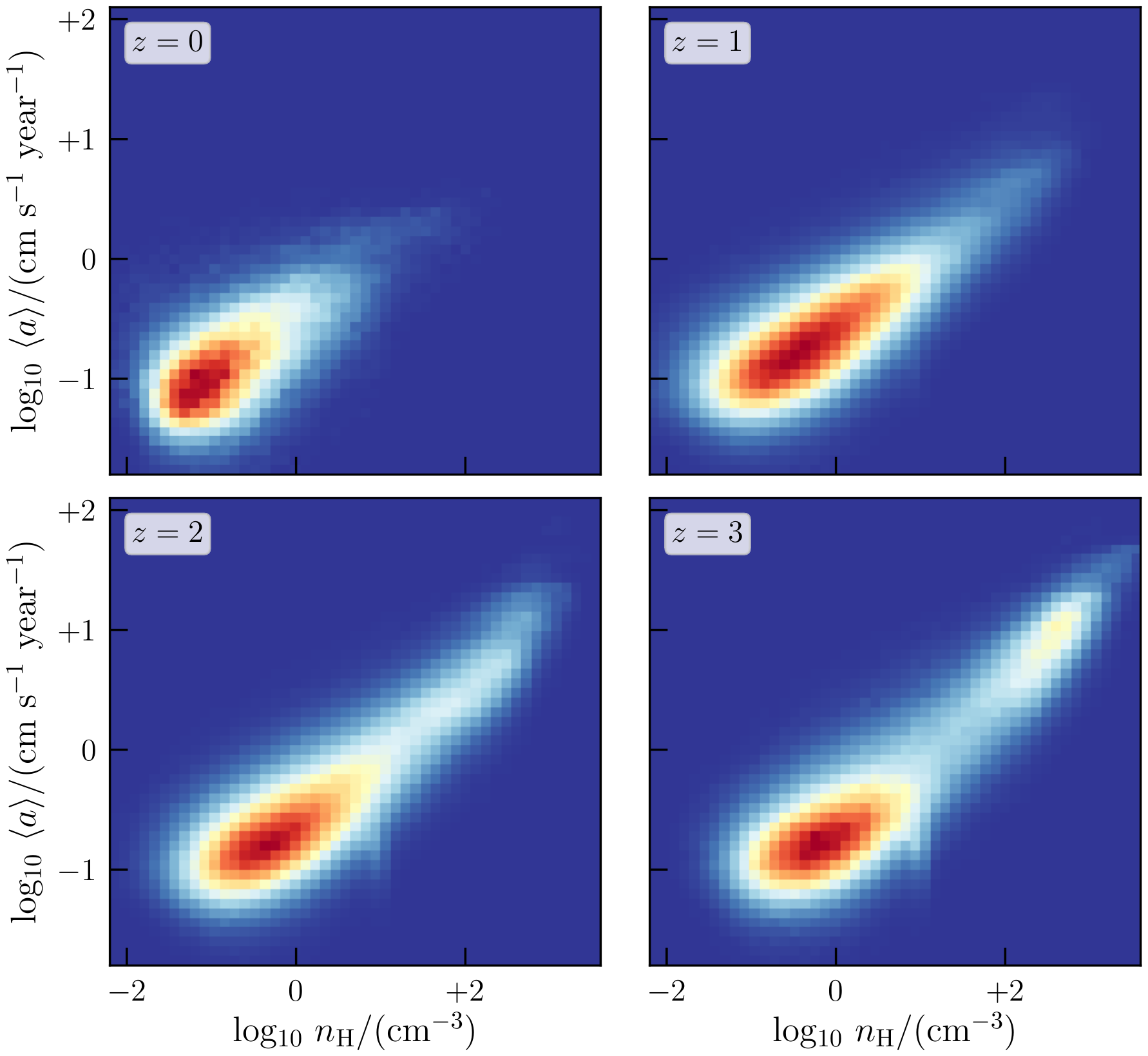}
    \caption{The dependence of the angle-averaged acceleration on the local gas volume density, for a selection of redshifts. The linear colourmap shows the frequency of occurrence (red shades indicate higher frequency). When the gas density $\log_{10}~n_{\rm H}/{\rm cm}^{-3}<0$, the gas particles are primarily influenced by the overall distribution of mass in a galaxy. At higher densities, the gravitational field is dominated by baryons that are in the immediate vicinity of the gas particle.}
    \label{fig:gravdens}
\end{figure*}

There are several notable features of these distributions. Across all redshifts, there is a peak of the angle-averaged acceleration around $2-4~{\rm mm~s}^{-1}~{\rm year}^{-1}$, which is comparable to, or somewhat larger than, the expected cosmological signal. The peak of the distribution is also similar in magnitude to the acceleration of the solar system barycentre\footnote{For consistency with the simulation values, we plot the angle-averaged acceleration of the solar system barycentre.} \citep{TitKra18}. The distributions also show an asymmetric tail to high $\langle a \rangle$ values. This feature is more prominent in the higher redshift snapshots. We found that the peculiar accelerations of gas particles near the peak of the $\langle a \rangle$ distributions occur when the local gravitational field is dominated by the overall distribution of matter in galaxies; the tail to high $\langle a \rangle$ occurs when the gas becomes self-gravitating (i.e. the gravitational field is dominated by a local concentration of baryons).\footnote{We confirmed that the acceleration experienced by the gas particles shown in Figure~\ref{fig:sim} is not dominated by a single or a small number of particles. Thus, the results are not affected by the particle nature of the simulations.}
This can be seen in Figure~\ref{fig:gravdens}, where we show a correlation between the local gas density and the strength of the gravitational field. We note that the acceleration in this regime scales as $a \sim v/t_{\rm dyn}\propto \sqrt{n_{\rm H}}$, where $t_{\rm dyn} = (G\,\rho)^{-1/2}$ is the dynamical time.

Thus, if the molecular or \HI~21~cm absorption line profiles are dominated by small clumps, the peculiar acceleration of these clumps may dominate over the cosmological signal by up to $\sim2$ orders of magnitude. However, given that the physical conditions of the EAGLE simulation are not resolved in this limit, future dedicated simulations \citep[e.g.][]{KimOstKim14} are required to assess the acceleration of individual absorption lines in star-forming regions. In the event that the `average' absorption line profile is representative of the orbital motion of the gas in the galaxy, we agree with previous studies that conclude the peculiar acceleration of gas in galaxies is comparable to or somewhat larger than the cosmological signal across all redshift intervals considered here. On a more positive note, if the asymmetric tail of the acceleration distributions shown in Figure~\ref{fig:sim} are realized in galaxies at $z\sim3$, it may be possible to secure a measurement of this galactic acceleration signal with ALMA (see e.g. Figure~\ref{fig:chanspace}).

\subsection{The peculiar acceleration of the \Lya\ forest}
\label{sec:lyaforest}

Almost every hydrogen atom in the Universe at $z\sim3$ is ionized. Nevertheless, there exists a large number of mostly ionized pockets of gas in the intergalactic medium (IGM) that contain low levels of neutral hydrogen. Empirically, their high incidence gave rise to the idea that these clouds are in predominantly low density regions, far away from galaxies \citep{Sar80}. This general picture is borne out of cosmological hydrodynamic simulations \citep[see][for an overview]{Mei09}, and intuitively, we might expect the \Lya\ forest to exhibit low peculiar accelerations \citep{Loe98,Lis08} since the lines arise far away from large overdensities of matter. Furthermore the observed sizes of the absorbers, based on a comparison of multiply-imaged or nearby quasar sightlines, indicates that low \HI\ column density, \NHI, absorbers trace large scale structures \citep{Sme92,Bec94,Sme95,Fan96,DOd06} whose motions are dominated by the Hubble flow \citep{Rau05}. For these reasons, the \Lya\ forest is considered an excellent tool to secure a measurement of the redshift drift.

Typically, the simulated IGM is studied by casting light rays through successive simulation outputs to simulate a quasar spectrum \citep{The98,Sch03,Bir15,Hum17,RotArt18}, which is then fit with Voigt profiles to construct a list of \HI\ absorbers and their properties. However, for the purpose of this work, we can use simple relations between the \HI\ column density and the gas volume density (or overdensity) for optically thin gas clouds \citep[][see also, \citealt{Dav10}]{Sch01}. Specifically, we convert the properties of every gas particle in the simulation output into an estimate of the \HI\ column density using the following relation:

\begin{figure*}
	\includegraphics[width=\textwidth]{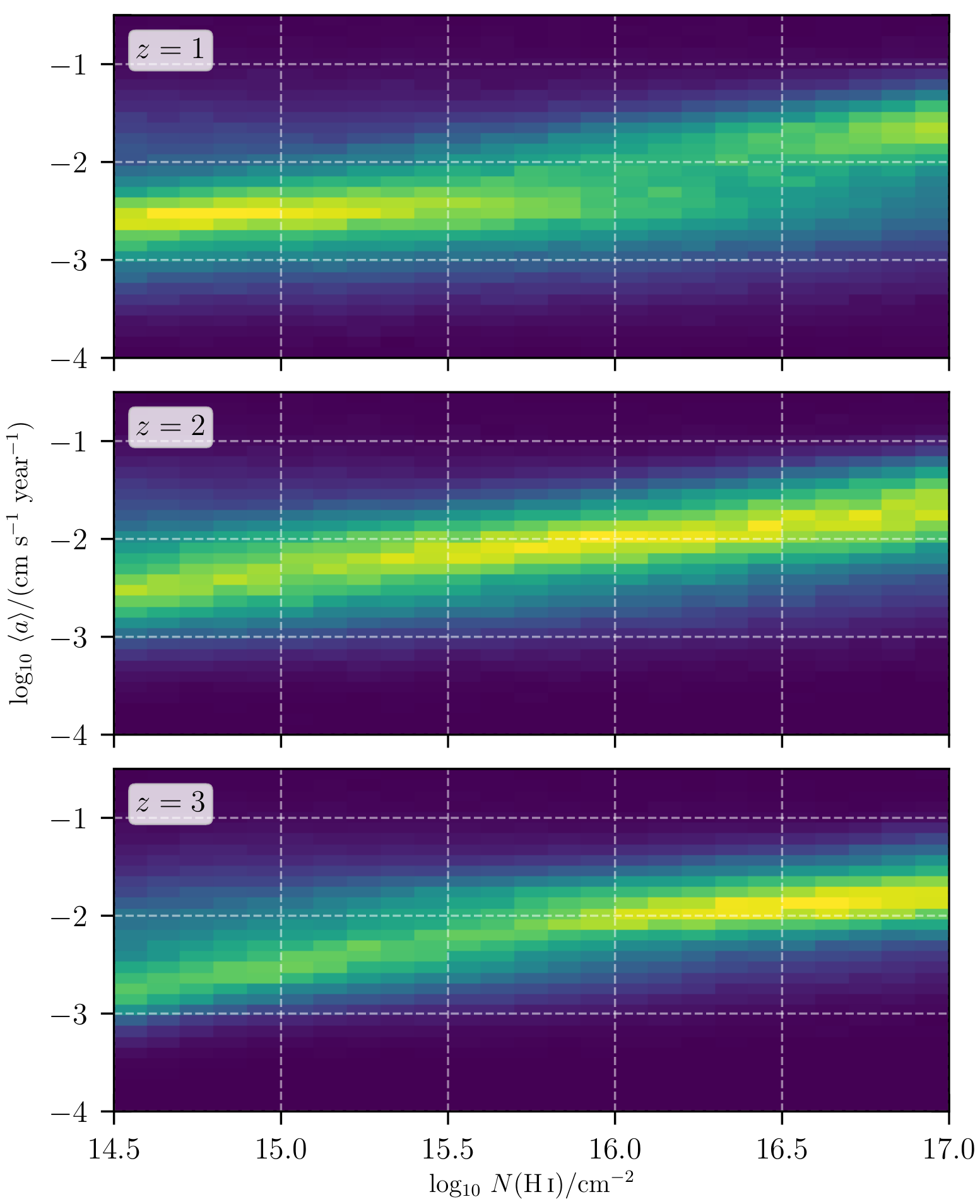}
    \caption{The angle-averaged acceleration of \Lya\ forest clouds as a function of the \HI\ column density. The top, middle and bottom panels show the 2D distributions at redshift $z=1, 2, 3$, respectively. The linear colourmap shows the frequency of occurrence (yellow shades indicate higher frequency). Every \HI\ bin contains 10,000 determinations of the peculiar acceleration; the brighter regions correspond to \HI\ column densities that exhibit a relatively narrower distribution of peculiar accelerations. Note that lower \HI\ column density absorption line systems typically exhibit lower peculiar acceleration.}
    \label{fig:simHI}
\end{figure*}

\begin{equation}
\label{eqn:nhi}
N({\rm H\,\textsc{i}}) = \frac{2.3\times10^{13}}{\Gamma_{\rm H\,\textsc{i}}} \bigg(\frac{n_{\rm H}}{10^{-5}~{\rm cm}^{-3}}\bigg)^{1.5}\bigg(\frac{T}{10^{4}\,{\rm K}}\bigg)^{-0.26}~\bigg(\frac{f_{\rm b}}{0.17}\bigg)^{0.5}
\end{equation}
where we use a baryon fraction $f_{\rm b}=0.1867$ \citep{Pla18}, and an \HI\ photoionization rate
$\Gamma_{\rm H\,\textsc{i}}/10^{-12}~{\rm s}^{-1} = 0.38,~0.94,~0.82$ at redshift $z=1,~2,~3,$ respectively \citep{HarMad12}.
We then apply a density and temperature cut ($n_{\rm H}<0.005~{\rm cm}^{-3}$ and $T<30,000$~K) to specifically select \HI\ absorption systems in the range $14.5<\log~N({\rm H\,\textsc{i}})/{\rm cm}^{-3}<17.0$. From this subset of particles, we randomly select 10,000 `absorbers' in the above column density range, in steps of 0.1~dex. When calculating the gravitational field, we include all particles within a radius of 1.5~pMpc of the test particle.\footnote{There was no noticeable change if we instead used a radius of 0.3~pMpc to compute the gravitational field.}

The angle-averaged distribution of the acceleration experienced by the absorbers in this column density range at redshift $z = 1,~2,~3$ are shown in Figure~\ref{fig:sim} as the thick yellow, blue, and red curves, respectively. Note that these angle-averaged acceleration distributions are not weighted by the incidence rate of absorbers; all absorbers are given equal weight. The \Lya\ forest distributions shown in Figure~\ref{fig:sim} exhibit a similar shape across the redshift interval considered here, with a peak around $0.01~{\rm cm~s}^{-1}~{\rm year}^{-1}$ --- more than an order of magnitude below the cosmological signal at $z\simeq3$.

Taking the acceleration distribution one step further, given the relation between overdensity and \NHI\ used here (see Equation~\ref{eqn:nhi}), it is reasonable to expect that \NHI\ depends on the local gravitational field if the particle gas density is a reflection of the local distribution of matter. These distributions are displayed in Figure~\ref{fig:simHI}, where yellow shades indicate more likely values of the angle-averaged acceleration. Across all redshifts considered, we find that low \NHI\ absorbers ($\log~N({\rm H\,\textsc{i}})/{\rm cm}^{-2}\sim14.5$) are less affected by peculiar acceleration than partial Lyman limit systems ($\log~N({\rm H\,\textsc{i}})/{\rm cm}^{-2}\sim17.0$).

The peculiar acceleration of the \Lya\ forest was previously explored by \citet{Lis08} using cosmological simulations. These authors estimated the peculiar acceleration by dividing the velocity of a particle by its dynamical time, $t_{\rm dyn}=(G\rho)^{-1/2}$.
As shown in Figure~\ref{fig:simHI}, we find that the peculiar acceleration depends gently on the \HI\ column density, with an approximate form, $a\propto N({\rm H\,\textsc{i}})^{1/3}$. Combined with Equation~\ref{eqn:nhi}, this suggests that $a\propto \sqrt{n_{\rm H}}$, which is identical to the result discussed in Section~\ref{sec:simcnm} and shown in Figure~\ref{fig:gravdens}. This lends support to the approximate form of the peculiar acceleration proposed by \citet{Lis08}.

In \citet{Lis08}, the peak of the peculiar acceleration distribution of the \Lya\ forest at $z\sim3$ is $a_{\rm pec}\simeq0.0003~{\rm cm~s}^{-1}~{\rm year}^{-1}$, approximately two orders of magnitude below the signal that we estimate here. However, note that in our work, we have uniformly weighted the peculiar acceleration of gas in the \HI\ column density range $14.5<\log~N({\rm H\,\textsc{i}})/{\rm cm}^{-3}<17.0$. If we instead weighted the peculiar acceleration by the \HI\ column density distribution function, we would significantly reduce the peak of the \Lya\ forest distribution, since lower \HI\ column density systems are considerably more numerous, and tend to exhibit lower peculiar acceleration. We therefore find a good general agreement between our simulation results and those of \citet{Lis08}.

\begin{figure*}
	\includegraphics[width=\textwidth]{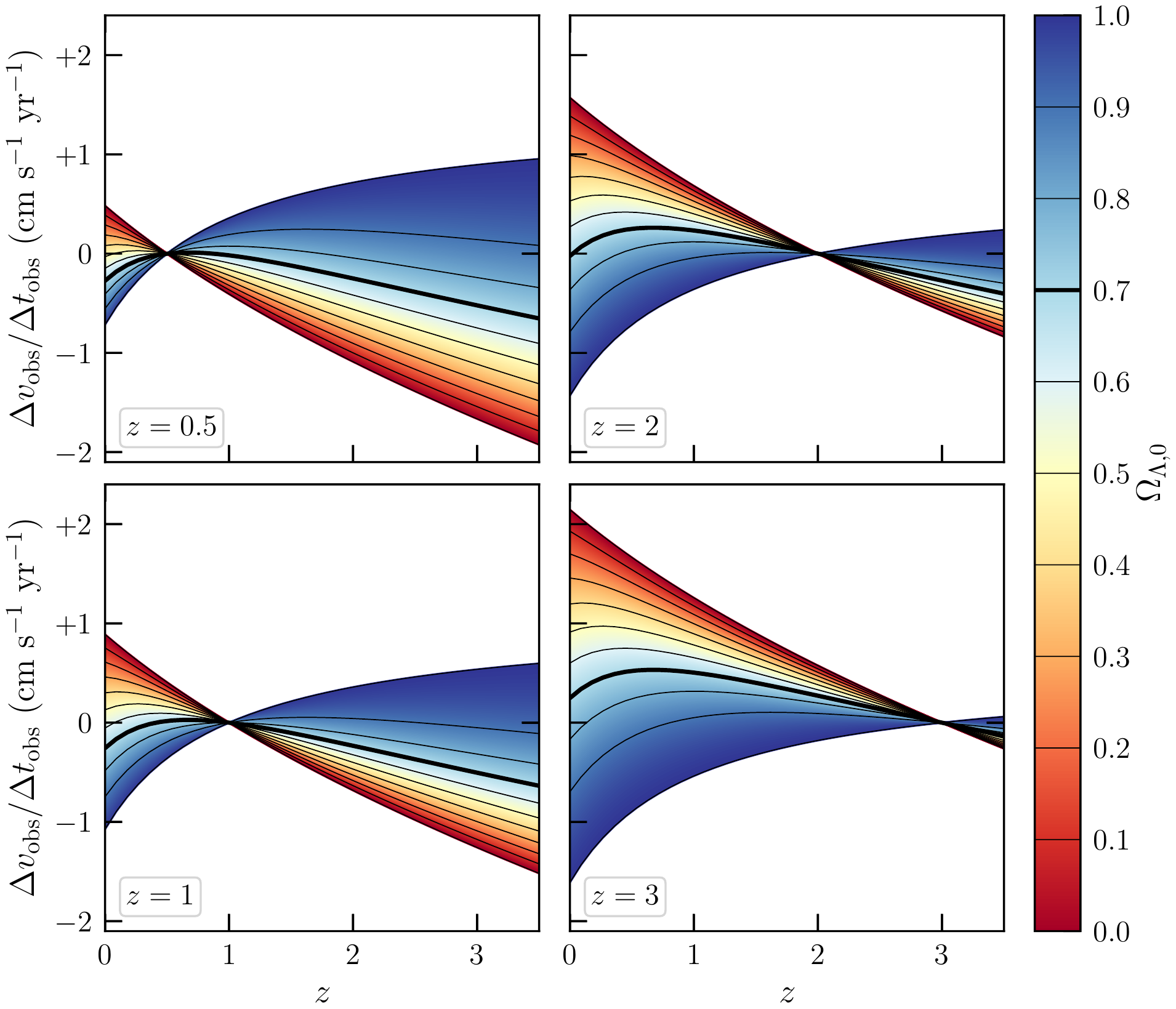}
    \caption{Similar to the bottom panel of Figure~\ref{fig:accel}, but instead shows the cosmic velocity drift of cosmological sources relative to another cosmological source at $z=0.5, 1, 2, 3$ (the comparison redshift is labelled on each panel).}
    \label{fig:reference}
\end{figure*}

In summary, our analysis suggests that a spectroscopic monitoring campaign of a single low \NHI\ absorber ($\log~N({\rm H\,\textsc{i}})/{\rm cm}^{-2}\sim14.5$) at redshift $z\sim3$ will likely exhibit a redshift drift that is dominated by the cosmological signal.

\section{Measuring the moving forest}
\label{sec:obs}

Given the favourable properties of the \Lya\ forest, as discussed in Section~\ref{sec:lyaforest}, in this section we explore some of the currently available data of the \Lya\ forest to place a limit on the cosmic redshift drift. Before doing so, it is instructive to first estimate the currently attainable accuracy of the redshift drift, given the \emph{absolute} accuracy of the wavelength calibration in optical echelle spectrographs that employ a ThAr calibration procedure. \citet{WhiMur15} estimate typical shifts of up to $\sim500~{\rm m~s}^{-1}$ (see also, \citealt{Rah13}). Given the time baseline of typical archival data ($\sim10$~years), we should expect that we can measure absolute velocity shifts down to a limiting value of no better than $\sim50~{\rm m~s}^{-1}~{\rm year}^{-1}$, which is still an order of magnitude poorer precision than that currently attained using heterodyne techniques in the radio wavelength regime \citep{Dar12}.

In principle, it should be possible to push past this absolute limit by instead measuring the \emph{relative} velocity shift between two unrelated sources that both move with the Hubble flow. For example, the \MgII\ absorption line doublet of a gas cloud at redshift $z_{1}\simeq0.7$ will be imprinted near the \Lya\ absorption line of an unrelated gas cloud at redshift $z_{2}\simeq2.9$. Provided that both gas clouds are known to arise from the \Lya\ forest (i.e. all absorption lines arise from gas clouds with an \HI\ column density $\log~N({\rm H\,\textsc{i}})/{\rm cm}^{-2}\lesssim17.0$), the drift of their relative redshift will have the following convenient properties:

\begin{enumerate}
    \item The \emph{absolute} wavelength calibration does not need to be known; only the \emph{relative} calibration between the two epochs needs to be determined. This can be achieved using the `\Lya\ cell' method described below (Section~\ref{sec:multiline}).
    \item For favourable redshift combinations (such as the \MgII+\Lya\ example given above), the signal of the cosmic redshift drift can be roughly doubled (in the case of the Standard Model), due to the change of sign of the cosmological redshift drift at low and high redshift (see dark solid line in the bottom panel of Figure~\ref{fig:accel}).
    \item Since we are performing a relative comparison of two cosmological sources along the same line-of-sight, the peculiar acceleration of the observer (i.e. the acceleration of the solar system barycentre around the Milky Way) does not need to be known. This correction can, in any case, be mitigated by observing quasars that are orthogonal to the direction of the barycentre acceleration.
    \item The relative barycentric velocity correction (i.e. the motion of the Earth around the solar system barycentre) between the two epochs will be minimised. Note, there is a small net barycentric correction that may need to be applied due to the curvature of the barycentric correction over the duration of each exposure \citep{Tro19}.
\end{enumerate}

Also, recall that the peculiar acceleration of both sources is expected to be subdominant, provided that both sources are associated with the \Lya\ forest (see Section~\ref{sec:lyaforest}); therefore, only a few combinations of favourable lines are needed to secure a direct measurement of the redshift drift. One of the obvious shortcomings of using a relative wavelength scale is that we must restrict ourselves to a small handful of wavelength regions; an absolutely calibrated wavelength scale, on the other hand, can utilize the full \Lya\ forest to measure the redshift drift, thereby greatly increasing the statistics. Figure~\ref{fig:reference} illustrates the sensitivity of the redshift drift to the parameters of a flat \lcdm\ model for a comparison source that is at redshift $z=0.5, 1, 2, 3$.

\subsection{The Ly$\alpha$ cell method}
\label{sec:multiline}

We now discuss a new technique that utilizes the many lines of the \Lya\ forest as the equivalent of a \Lya\ `iodine cell' (hereafter, referred to as the `\Lya\ cell') across multiple observing epochs. With this approach, the centroids of successive \Lya\ forest lines can be matched across epochs, allowing the redshift drift of an intervening metal line or high order Lyman series line to be observed to shift relative to the \Lya\ forest lines.

Taking the first observing epoch to represent the truth,
the redshift of absorption lines from future epochs can be described by a small correction (assumed to be linear for the present illustration) to the wavelengths:
\begin{equation}
    z_{2} = z_{1} + \alpha_{2}\,\lambda_{2} + \beta_{2} + \Delta z_{\rm obs}
\end{equation}
where $z_{1}$ and $z_{2}$ are the measured redshifts of an absorption line at the first and second observing epochs respectively, $\alpha_{2}$ and $\beta_{2}$ represent a stretch and shift that are applied to the uncorrected wavelength solution at the second epoch ($\lambda_{2}$) at the location of the absorption line, and $\Delta z_{\rm obs}$ corresponds to the relative cosmological drift between the two epochs. Now, let us consider an intervening absorber at redshift $z_{1i}$, relative to a reference (i.e. \Lya) line $z_{1a}$. Noting that $\lambda_{2a}=(1+z_{2a})\lambda_{0a}$ and $\lambda_{2i}=(1+z_{2i})\lambda_{0i}$, where $\lambda_{0a}$ and $\lambda_{0i}$ are the rest frame line wavelengths of \Lya\ and the intervening absorption line, we have the following expression to deduce the redshift of the intervening line at future epochs:
\begin{equation}
\label{eqn:z2i}
    z_{2i} = \frac{(z_{1i}-z_{1a}) + \alpha_{2}(\lambda_{0i}-\lambda_{0a}) + Q\,\Delta t_{\rm obs} + z_{2a}(1-\alpha_{2}\lambda_{0a})}{1-\alpha_{2}\lambda_{0i}}
\end{equation}
where $Q\,\Delta t_{\rm obs} \equiv (\dot{z}_{i}-\dot{z}_{a})\,\Delta t_{\rm obs} = \Delta z_{{\rm obs},i}-\Delta z_{{\rm obs},a}$. Therefore, for every new epoch, the only free parameters are the scale $\alpha_{2}$, and the redshift of a reference line $z_{2a}$. Note, the shift correction ($\beta_{2}$) does not appear explicitly in Equation~\ref{eqn:z2i}; the shift correction is accounted for with the parameter $z_{2a}$. One can write a similar expression for the lines that comprise the `\Lya\ cell':
\begin{equation}
\label{eqn:cell}
    z_{2b} = z_{2a} + \frac{z_{1b}-z_{1a}}{1-\alpha_{2}\lambda_{0a}}
\end{equation}
Note, this is identical to Equation~\ref{eqn:z2i}, with the replacement $\lambda_{0i}\to\lambda_{0b}=\lambda_{0a}$.
Equation~\ref{eqn:cell} encapsulates the advantage of this technique; having many lines in a \Lya\ cell will allow the $\alpha_{2}$ stretch correction to be pinned down. Thus, with the parameters of the \Lya\ cell constrained, the only free parameter is the cosmological shift between the cell and the intervening line, $Q\equiv\dot{z}_{i}-\dot{z}_{a}$ (refer to Equation~\ref{eqn:dzdt}).

We also note that, if an observation can be acquired with an accurate absolute wavelength calibration (e.g. such as that made possible with VLT+ESPRESSO, or with the instruments planned for the future generation of extremely large telescopes), then one can use the \Lya\ forest lines as an iodine cell to accurately re-calibrate spectra acquired decades earlier, for example, the data acquired with the current generation of $8-10$~m class telescopes. This approach may permit a measurement of the redshift drift using the intervening metal absorption lines (relative to the \Lya\ forest).

Finally, it is emphasized that the \Lya\ cell technique can in principle improve the wavelength calibration whenever the small corrections to the wavelength solution are nearly linear over the wavelength range of interest. Thus, even in the future when sufficiently stable instrumentation is able to secure a direct measurement of the redshift drift, the \Lya\ cell technique may allow us to push beyond the absolute wavelength calibration of the available instrumentation by instead securing a \emph{relative} measurement of the redshift drift. However, this will require us to pre-plan the brightest quasars that contain suitable combinations of metal+\Lya\ lines. We also speculate that a closely spaced doublet (e.g. \CIV) may be used in combination with a metal line (e.g. \MgII) to achieve higher accuracy, due to the greater structure and narrow shape of the metal line profiles compared with the \Lya\ forest lines, which tend to be broader.

\subsection{Acceleration measures toward HS\,1700$+$6416}

To demonstrate the utility of the \Lya\ cell technique, we apply it to several example intervening metal absorption lines along the line-of-sight to the high redshift ($z_{\rm qso}=2.7348$) quasar HS\,1700$+$6416. This bright $V=16.1$ quasar has been observed multiple times over a decade with the Keck telescope, and the data products were uniformly reduced and made publicly available as part of the Keck Observatory Database of Ionized Absorption toward Quasars (KODIAQ)\footnote{KODIAQ data products are available from:\newline \url{https://koa.ipac.caltech.edu/cgi-bin/KODIAQ/nph-kodiaqStartup}} project \citep{OMe15,OMe17}.

HS\,1700$+$6416 is a high redshift quasar made famous by its nearly transparent sightline; the highest \HI\ column density system along this line-of-sight is $\log~N({\rm H\,\textsc{i}})/{\rm cm}^{-2}\simeq17.0$ \citep{Fec06}. This suggests that the intersected \Lya\ forest lines are almost entirely associated with relatively underdense environments that have low peculiar acceleration (see Section~\ref{sec:lyaforest}). Publicly available data of HS\,1700$+$6416 from the KODIAQ project span an observation baseline of over 12 years, from 1996 May 22 (MJD=50226.56) to 2008 September 25 (MJD=54734.28), and include ten independent spectra. Having not been observed now for over 10 years, this sightline offers the immediate advantage of being able to readily improve the redshift drift constraints by a factor of $\sim2$. Furthermore, one of the most important advantages of this sightline is that the absorption lines are well-categorized \citep[see e.g.][]{Tri97,Fec06}; the observed wavelengths of all intervening metal absorption lines are known, thereby giving us confidence that the combination of \Lya\ forest lines chosen for study here are uncontaminated by unrelated metal absorption lines.

\begin{figure*}
	\includegraphics[width=\textwidth]{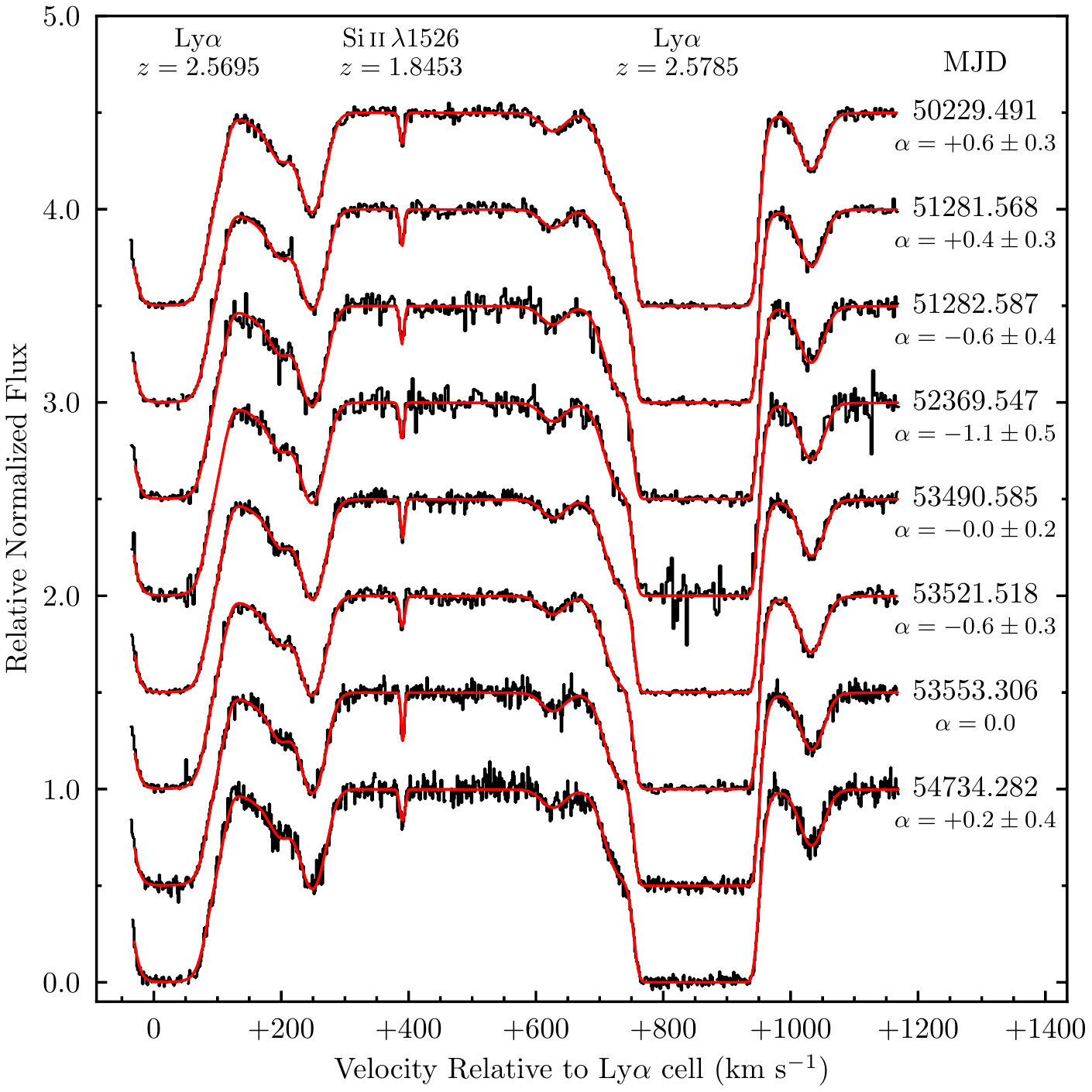}
    \caption{Multi-epoch continuum normalized spectra of the quasar HS\,1700$+$6416 (black histograms; MJD of each spectrum is indicated on the right of the panel) overlaid with the overall best-fitting model (red line). Each consecutive epoch is offset in the vertical direction by $+0.5$ for clarity. The stretch parameter ($\alpha$) of each epoch is listed below the MJD, and is in units of $10^{-6}\,$\AA$^{-1}$. The epoch labelled $\alpha=0.0$ corresponds to the reference epoch; the stretch parameter of each epoch therefore corresponds to the wavelength stretch \emph{relative} to the reference epoch. Data that are affected by cosmic rays and/or defects are not shown (see text). \SiII\,$\lambda1526$ absorption (at $v\simeq+390~{\rm km~s}^{-1}$) at redshift $z\simeq1.8453$ is located between numerous \Lya\ absorbing clouds at redshift $z\simeq2.57$. Note, the line profile structure of each individual line is the same for all epochs; the apparent difference in the line profiles is due to the various instrumental broadening functions that were adopted by different observers. For reference, the expected relative line shift between the \Lya\ cell and the \SiII\ absorption is $\dot{v}\simeq0.19~{\rm cm~s}^{-1}~{\rm year}^{-1}$, assuming the standard cosmological model. The model shown corresponds to a reduced chi-squared $\chi^{2}_{\rm red}=1.304$.}
    \label{fig:fitsA}
\end{figure*}

\begin{figure*}
	\includegraphics[width=\textwidth]{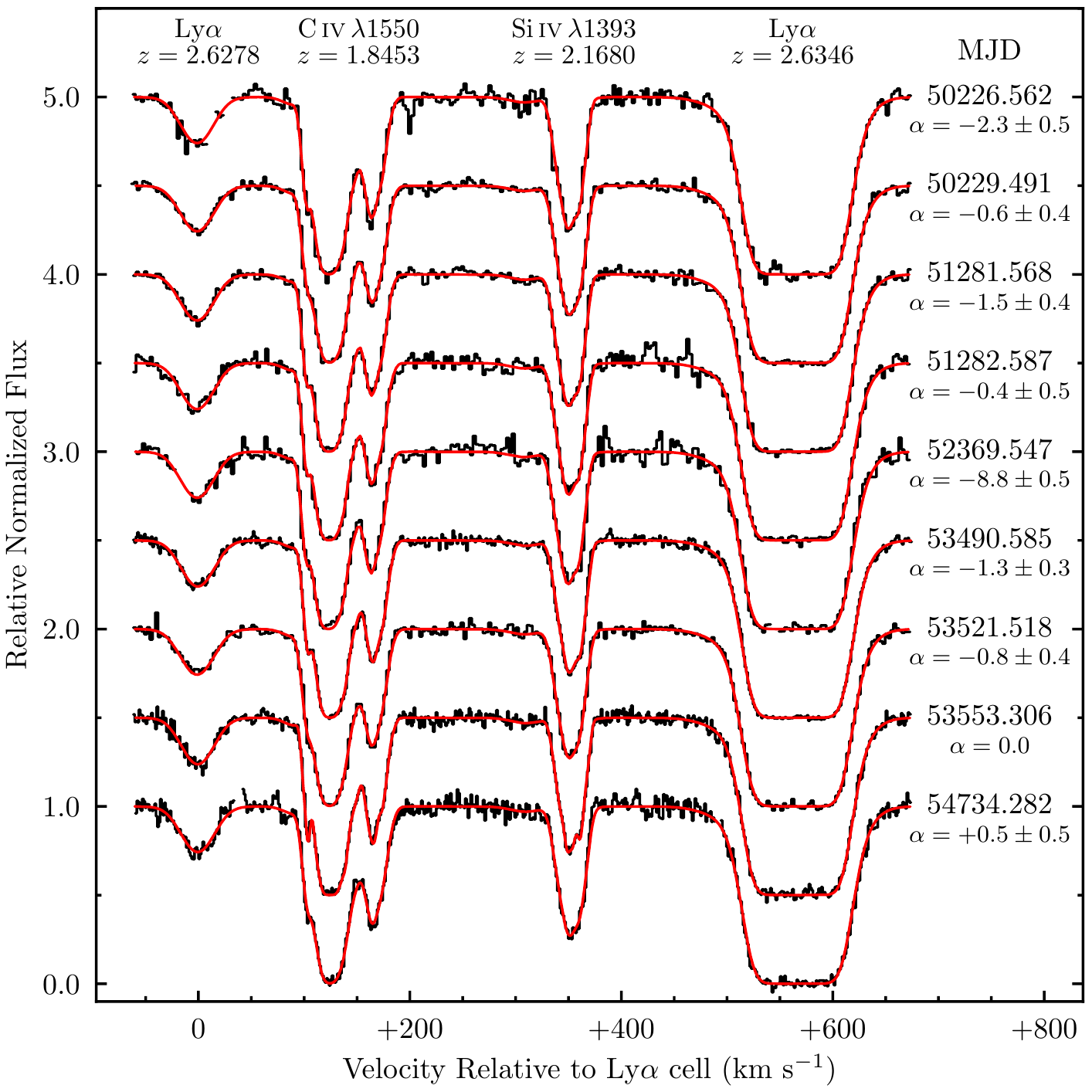}
    \caption{Same as Figure~\ref{fig:fitsA}, except there are two unrelated metal absorption lines (\CIV\,$\lambda1550$ at $z\simeq1.8453$ and \SiIV\,$\lambda1393$ at $z\simeq2.1680$) located between two \Lya\ absorbing clouds (at $v\simeq0~{\rm km~s}^{-1}$ and $v\simeq+570~{\rm km~s}^{-1}$) at redshift $z\simeq2.63$. For reference, the expected relative line shift between the \Lya\ cell and the \CIV\ and \SiIV\ absorption is $\dot{v}\simeq0.21~{\rm cm~s}^{-1}~{\rm year}^{-1}$ and $\dot{v}\simeq0.12~{\rm cm~s}^{-1}~{\rm year}^{-1}$, respectively, assuming the standard cosmological model. The model shown corresponds to a reduced chi-squared $\chi^{2}_{\rm red}=1.418$.}
    \label{fig:fitsB}
\end{figure*}

\begin{figure*}
	\includegraphics[width=\textwidth]{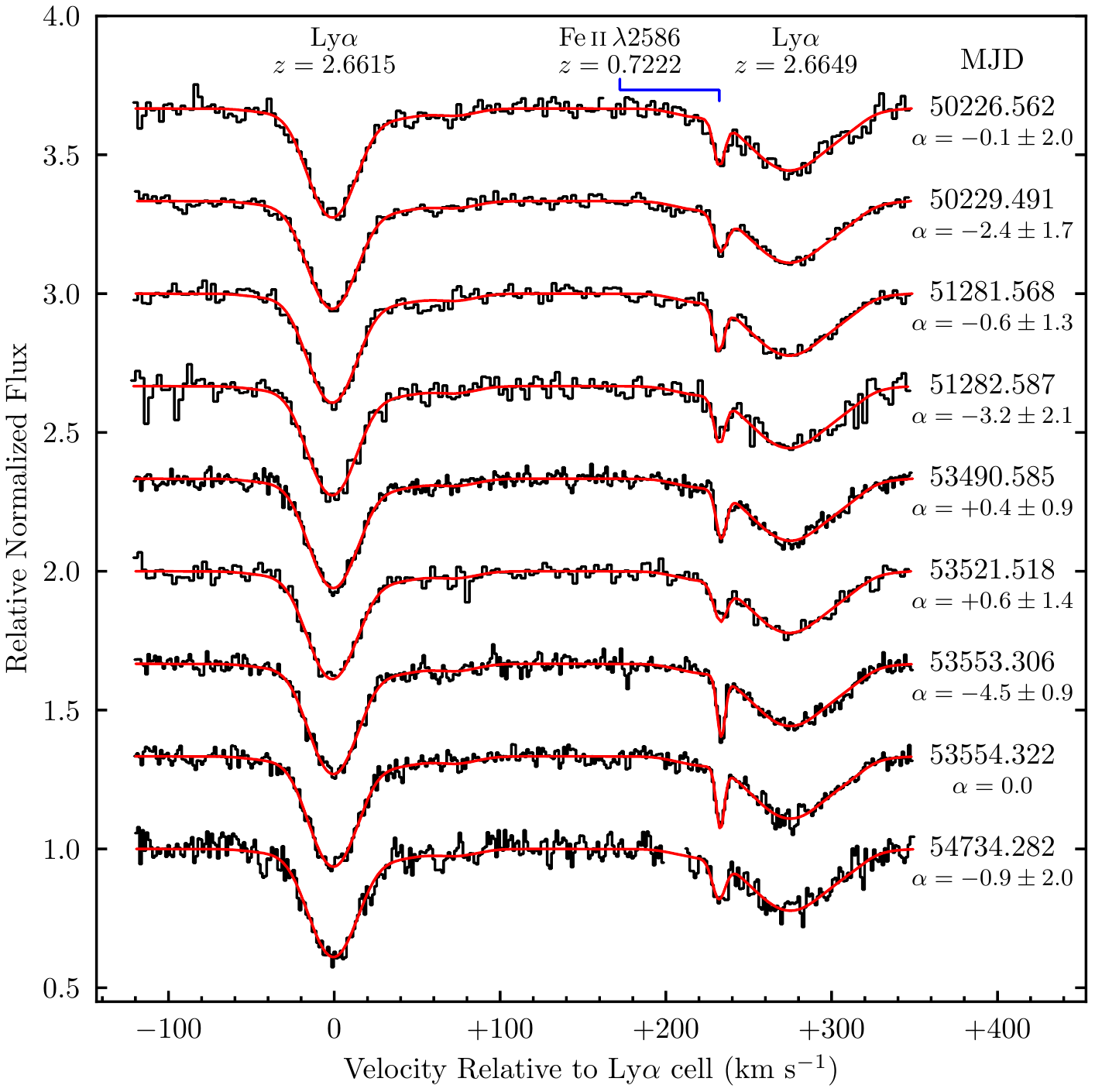}
    \caption{Same as Figure~\ref{fig:fitsA}, except for \FeII\,$\lambda2586$ absorption (at $v\simeq+230~{\rm km~s}^{-1}$) at redshift $z\simeq0.7222$ is located between two \Lya\ absorbing clouds (at $v\simeq0~{\rm km~s}^{-1}$ and $v\simeq+270~{\rm km~s}^{-1}$) at redshift $z\simeq2.66$. For reference, the expected relative line shift between the \Lya\ cell and the \FeII\ absorption is $\dot{v}\simeq0.44~{\rm cm~s}^{-1}~{\rm year}^{-1}$, assuming the standard cosmological model. The model shown corresponds to a reduced chi-squared $\chi^{2}_{\rm red}=1.189$.}
    \label{fig:fitsC}
\end{figure*}

We use the Absorption LIne Software (\textsc{alis}; \citealt{Coo14})\footnote{\textsc{alis} is publicly available on GitHub:\newline \url{https://github.com/rcooke-ast/ALIS}} to fit all of the available high spectral resolution ($v_{\rm FWHM}\lesssim8~{\rm km~s}^{-1}$) data of HS\,1700$+$6416. \textsc{alis} uses a modified version of the \textsc{mpfit} software \citep{Mar09} to minimize the $\chi^{2}$ statistic. We simultaneously fit for the quasar continuum and intervening absorption lines to ensure that any uncertainty in the continuum placement across multiple epochs is accounted for in the quoted error budget. We use a low order Legendre polynomial for the quasar continuum (typically of order 4), and we use a series of Voigt profiles to model the absorption lines. Each Voigt profile is described by a column density, redshift, and Doppler parameter. For a given absorption line, we tie the parameters of the Voigt profile to be the same across all epochs. This constraint assumes that the cloud structure is invariant with time, and that the quasar sightline probes the same gas at all epochs.

\begin{table}
	\centering
	\caption{Acceleration measures towards HS\,$1700+6416$}
	\label{tab:Qresults}
	\begin{tabular}{cccccc}
		\hline
		Reference & $z_{\rm ref}$ & Intervening & $z_{\rm int}$ & $Q$ & $Q_{\rm raw}$\\
		Ion &  & Ion &  & $(10^{-7}~{\rm year}^{-1})$ & $(10^{-7}~{\rm year}^{-1})$\\
		\hline
		\Lya & $2.5757$ & \SiII & $1.848$ & $5.3\pm3.4$ & $7.0\pm2.8$ \\
		\Lya & $2.6317$ & \CIV & $1.848$ & $5.1\pm3.5$ & $-1.22\pm0.50$ \\
		\Lya & $2.6317$ & \SiIV & $2.167$ & $3.9\pm2.1$ & $0.1\pm1.3$ \\
		\Lya & $2.6638$ & \FeII & $0.721$ & $1.8\pm5.5$ & $10.0\pm2.4$ \\
		\hline
	\end{tabular}
\end{table}

The model is generated on a wavelength scale that is a factor of $\sim30$ higher sampling than the native pixel resolution, to account for the nonlinear profile structure within a pixel, which is particularly important near the core of the absorption line.\footnote{We confirmed that the results are unchanged if we use a subpixellation scale that is a factor of 300 below the native pixel resolution.} The model is then convolved with the nominal instrument resolution.\footnote{Derived from the widths of ThAr calibration lines.} We allow the measured wavelength scale of all epochs to be adjusted relative to the reference epoch\footnote{The reference epoch was chosen arbitrarily. This choice does not impact our analysis of the \emph{relative} adjustments of the wavelength scale.} (i.e. two free parameters for every epoch, corresponding to $\alpha_{2}$ and $z_{2a}$ in Equation~\ref{eqn:z2i}); as described in Section~\ref{sec:multiline}, this ensures that the \Lya\ cell is matched across all epochs and allows us to measure the drift of the intervening metal lines relative to the \Lya\ lines. Finally, we introduce a single free parameter, $Q$ (recall: $Q\equiv\dot{z}_{i}-\dot{z}_{a}$), which is the same for all epochs. As a final note, some data are affected by defects, mostly due to cosmic rays. Regions that are affected by cosmic rays were visually identified and masked during the fitting procedure; these pixels do not contribute to the model $\chi^{2}$.

For the purposes of this paper, we present four example measurements of the \Lya\ cell technique using the \Lya\ forest at $z\simeq2.55-2.65$, which is just lower than the redshift of the quasar ($z_{\rm qso}=2.7348$); this allows us to use the highest redshift \Lya\ forest lines available, and potentially maximizes the relative shift between the \Lya\ cell and the intervening lines. We note that, although there are ten independent spectra of this quasar available in the KODIAQ database, some of these spectra do not cover the \Lya\ cells that are considered here. Thus, our analysis is only based on eight or nine epochs of data. Model fits to these absorption line systems are shown in Figures~\ref{fig:fitsA}--\ref{fig:fitsC}. The measured $Q$ value of each individual system is listed in the penultimate column of Table~\ref{tab:Qresults}. For comparison, in the final column of this table, we also list the $Q$ value derived for the same model but applying no stretch correction ($Q_{\rm raw}$; note, a shift correction is still applied). In one case ($z_{\rm ref}=2.6638$), the $Q_{\rm raw}$ value deviates from zero by $\sim4\sigma$, while the corrected value is consistent with zero. All other measures of $Q$ are consistent with zero acceleration (to within $2\sigma$). We combine the results of all four measures of $Q$ to
derive a $2\sigma$ upper limit on the cosmological velocity drift, $\dot{v}_{\rm obs}<65~{\rm m~s}^{-1}~{\rm year}^{-1}$, which is still almost four orders of magnitude larger than the cosmological signal, but is nevertheless the best available limit based on optical \Lya\ forest spectra. The limit reported above is a factor of $\sim6$ lower precision than the $2\sigma$ limit reported by \citet[][$\dot{v}_{\rm obs}<10~{\rm m~s}^{-1}~{\rm year}^{-1}$]{Dar12}, which was based on absolute drift measurements of 10 absorption line systems.

In principle, the limit reported here could be improved substantially by analyzing all available multi-epoch data from the KODIAQ database (138/300 quasars), in addition to the Spectral QUasar Absorption Database (SQUAD) data that were acquired using the Ultraviolet and Visual Echelle Spectrograph (UVES) on the Very Large Telescope (113/467 quasars; \citealt{Mur19}). Assuming that, on average, each of these $\sim250$ quasars with multi-epoch data contain $\sim10$ \Lya\ cells, each delivering a precision of $\sim5\times10^{-7}~{\rm year}^{-1}$ (a typical value based on the four cells analyzed here), we can estimate the final expected precision on $Q$ to be $\Delta Q\approx10^{-8}~{\rm year}^{-1}$ ($\dot{v}_{\rm obs}\approx30~{\rm cm~s}^{-1}~{\rm year}^{-1}$), which is still $\sim2$ orders of magnitude away from the expected cosmological signal. Such an analysis would constitute a heroic effort for a likely limited scientific gain! To conclude more positively, the \Lya\ cell technique described here opens up the possibility -- using currently available data -- to measure the acceleration of gas clouds along the line-of-sight to quasars down to a precision of $\approx10~{\rm m~s}^{-1}~{\rm year}^{-1}$, given a sufficiently long time baseline ($\sim20$~years). Such an acceleration might be exhibited by gas clouds in the immediate vicinity of a quasar.

\section{Summary and Conclusions}
\label{sec:conc}

We introduce the ACCELERATION programme, which has
the defining goal to characterize and study the
properties of the most underdense regions of the
Universe. This paper focuses on the possibility
of measuring the redshift drift using the \Lya\
forest. The main conclusions of this paper can
be summarized as follows:\\

\noindent ~~(i) Using the EAGLE cosmological
hydrodynamic simulations, we have estimated the
peculiar acceleration of absorption line systems.
We find that gas clouds that are conducive to
star formation (such as those in either the cold
neutral or molecular phases) are likely to exhibit
a peculiar acceleration comparable to or somewhat
larger than the cosmological signal. The distribution
of these gas clouds exhibits a tail of systems with
high peculiar acceleration, particularly at $z\sim3$.
This tail is predominantly due to the high density
of baryons in the regions surrounding the star
forming gas. To secure a clean measurement of the
cosmic redshift drift using such gas clouds would
require in excess of $\sim1000$ sightlines to be
sensitive to the cosmological signal.

\smallskip

\noindent ~~(ii) Using the same simulations, we also
explore the peculiar acceleration of low density \Lya\
forest clouds, with \HI\ column densities $N({\rm H\,\textsc{i}})<10^{17}~{\rm cm}^{-2}$. We find that the
typical peculiar acceleration of the \Lya\ forest is
more than an order of magnitude below the cosmological
signal, making the \Lya\ forest a near ideal environment
to secure a direct measure of the cosmic expansion
history. This general conclusion is in good agreement
with previous works \citep{Loe98,Lis08}. We also find
a gentle correlation of increasing peculiar acceleration
with increasing \HI\ column density, which is borne out
of the relationship between the \HI\ column density and
overdensity \citep{Sch01,Dav10}. It is therefore apparent
that the abundant, low \HI\ column density absorption lines
of the \Lya\ forest offer the most reliable environment to
pin down the cosmic expansion history.

\smallskip

\noindent ~~(iii) We propose a new `\Lya\ cell' technique
to measure the \emph{relative} redshift drift between two
absorption lines (e.g. \Lya\ and \MgII) that are close in
terms of their observed wavelength but are well-separated
in redshift. This method offers a number of
convenient properties, including the mitigation of the observer
acceleration, and the ability to overcome problems with the
accuracy and stability of the wavelength calibration over multiple epochs. The obvious shortcoming of this technique is that a \Lya\ cell measurement is restricted to a small wavelength range around each cell; on the other hand, a spectrum that is accurately wavelength calibrated can utilize the entire \Lya\ forest.

\smallskip

\noindent ~~(iv) We apply the \Lya\ cell technique to four
suitable combinations of \Lya+metal absorption lines toward
the quasar HS\,1700$+$6416, based on multi-epoch data secured
over a time baseline of $\sim12$ years. The redshift drift of
all four combinations are consistent with zero after accounting
for a relative shift and stretch to the wavelength solution. The
joint constraint of these four measures yields a $2\sigma$ limit
on the cosmic redshift drift
$\dot{v}_{\rm obs}<65~{\rm m~s}^{-1}~{\rm year}^{-1}$.

\smallskip

\noindent ~~(v) Based on the currently available multi-epoch
observations of quasars, we estimate that there are insufficient
\Lya\ cells to make an informative measurement of the cosmic
expansion rate at present.

\smallskip

Future technologies, such as laser frequency combs and the next generation of large aperture telescopes, promise to deliver a reliable and stable wavelength calibration. Indeed, a facility that is dedicated to this cosmology experiment would be an important step towards realizing this fundamental measurement, as highlighted recently \citep{Eik19}. We speculate that a small handful of \Lya\ cells, observed with VLT+ESPRESSO and future telescope facilities, may help to reduce some of the instrumental and observational systematics. However, some pre-planning of the observed quasars will be required to ensure that the most desirable \Lya\ cells are observed.

\section*{Acknowledgements}
I am grateful to the referee, Joe Liske, for providing a detailed and supportive report that helped me to clarify several key points throughout the paper.
I also thank Alejandro Ben\'itez-Llambay and Tom Theuns for helpful discussions regarding the interrogation of the EAGLE simulations.
During this work, R.~J.~C. was supported by a
Royal Society University Research Fellowship.
R.~J.~C. acknowledges support from STFC (ST/L00075X/1, ST/P000541/1).
This work used the DiRAC Data Centric system at Durham University,
operated by the Institute for Computational Cosmology on behalf of the
STFC DiRAC HPC Facility (www.dirac.ac.uk). This equipment was funded
by BIS National E-infrastructure capital grant ST/K00042X/1, STFC capital
grant ST/H008519/1, and STFC DiRAC Operations grant ST/K003267/1
and Durham University. DiRAC is part of the National E-Infrastructure.
This research has made use of NASA's Astrophysics Data System.


\bsp	
\label{lastpage}
\end{document}